\documentclass{aa}
\usepackage[varg]{txfonts}
\usepackage{graphicx}
\usepackage{epstopdf}
\usepackage{bm}

\usepackage{natbib}
\bibpunct{(}{)}{;}{a}{}{,} % to follow the A&A style

\begin{document}

\title{Ly$\alpha$ irradiation of solid-state formamide}
\titlerunning{UV processing of formamide ice}

   \author{T. Suhasaria
          \inst{1, 2}
          \and
          V. Mennella\inst{1}
          }

   \institute{INAF-Osservatorio Astronomico di Capodimonte, Salita Moiariello 16, 80131, Naples, Italy\\
         \and
             Current address: Max Planck Institute f\"ur Astronomie, K\"onigstuhl 17, 69117, Heidelberg, Germany\\
             \email{suhasaria@mpia.de}
             }

\abstract
{Formamide (NH$_2$CHO), a potential prebiotic precursor, has been proposed to play an important role in the context of origin of life on our planet. It has been observed in different environments in space including the protostellar regions and comets. The abundance and stability of NH$_2$CHO in the early stages of star formation can be better understood by incorporating the formation and destruction data in the astrochemical models.}
{We carried out an experimental investigation to study the destruction of pure NH$_2$CHO ice at 12 K by the interaction of Ly$\alpha$ (121.6 nm) photons.}
{The UV photo destruction of NH$_2$CHO was studied using Fourier-transform infrared spectroscopy.}
{After UV processing, the intensity of NH$_2$CHO IR bands decreases and new bands corresponding to HCN, CO, NH$_4^+$ OCN$^-$, HNCO, and CO$_2$ appeared in the spectrum. Destruction and cumulative product formation cross-sections were derived.}
{The comparison of destruction rate derived from the cross-section in cold and dense molecular cloud for different energetic processing agents, reveals that UV photons induces an order of magnitude higher NH$_2$CHO destruction than cosmic rays, but three orders of magnitude lower than for H atoms.}


\maketitle

\section{Introduction}

Formamide (NH$_2$CHO) is one of the simplest molecule that contains the four primary biogenic elements and an amide bond. It is considered a promising precursor molecule in the context of origin of life on our planet \citep{Saladino2012}. Formamide chemistry in controlled terrestrial environment allowed the formation of essential biological life components such as amino acids, nucleobases, sugars, carboxylic acids \citep{Saladino2012, Ferus2015, Botta2018}. Moreover, it has been observed in many astrophysical environments such as galactic centres, star-forming regions and comets [e.g. \citep{Blake1986, Turner1991, Bisschop2007b, Mendoza2014, Lopez2015, Bockelee2000, Biver2014, Goesmann2015}]. In fact, mass spectrometric analysis by COmetary Sampling And Composition (COSAC) aboard Rosetta's Philae lander on comet 67P/Churyumov–Gerasimenko, revealed that NH$_2$CHO [1.8\% with respect to water (H$_2$O)] is the second most abundant molecule after H$_2$O \citep{Goesmann2015, altwegg2017}. Solid state NH$_2$CHO has also been identified tentatively in the interstellar medium by Infrared Space Observatory towards NGC 7538 IRS9 and W33A \citep{raunier2004a, Gibb2004}. 

Based on experimental and theoretical studies, NH$_2$CHO has been proposed to be formed in space by a series of ion-molecule and neutral-neutral reaction channels in the gas phase whereas solid NH$_2$CHO could be formed via processing of interstellar ices on dust grain surface by electrons, ions, ultraviolet (UV) photons, heat and H atoms \citep[references therein]{Lopez2019}. Once formed, NH$_2$CHO can undergo further processing both in the gas or solid phase to form simple and complex organic molecules. As a result, the abundance of NH$_2$CHO in a certain space environment will depend both on the formation and destruction channels. Therefore, understanding the destruction channels of NH$_2$CHO under energetic and non-energetic processing has been the subject of numerous studies. 

Quantum mechanical calculations on the thermal decomposition of NH$_2$CHO revealed three main reaction channels \citep{nguyen2011}. In decreasing order of kinetically favoured pathway, they are dehydration [H$_2$O loss to hydrogen cyanide (HCN)], decarbonylation [carbon monoxide (CO) loss to ammonia (NH$_3$)] and dehydrogenation [hydrogen (H$_2$) loss to isocyanic acid (HNCO)]. High energy laser spark induces the decomposition of NH$_2$CHO gas to HCN, CO, NH$_3$, carbon dioxide (CO$_2$), nitrous oxide (N$_2$O), hydroxylamine (HONH$_2$), and methanol (CH$_3$OH) which was identified by infrared (IR) spectroscopy \citep{ferus2011}. NH$_2$CHO decomposition in matrices also leads to similar products as observed in the gas phase reactions \citep{lundell1998, maier2000, duvernay2005}.
Only limited studies have been dedicated to investigate the destruction of NH$_2$CHO in the condensed phase. Ion irradiation experiments with 200 keV H$^+$ were performed on NH$_2$CHO films deposited on inert silicon and olivine substrate \citep{Brucato2006a,brucato2006b}. IR analysis showed the formation of CO, CO$_2$, NH$_3$, N$_2$O, HCN, cyanide ion (CN$^-$), ammonium ion (NH$_4^+$), isocyanate ion (OCN$^{-}$) and isocyanic acid (HNCO) on silicon whereas all the other products except CN$^-$ and NH$_3$ were formed on the olivine surface. In a different set of experiments, pure NH$_2$CHO films show no significant degradation upon processing by UV-enhanced xenon lamp, whereas oxide minerals titanium dioxide (TiO$_2$) and spinel (MgAl$_2$O$_4$) used as substrate induce a gradual but minimal degradation \citep{Corazzi2020}. On the other hand, Lyman (Ly)$\alpha$ photons and 1 keV electrons bombardment of amorphous or crystalline NH$_2$CHO film deposited on silica (SiO$_2$) nanopowder at 70 K revealed only OCN$^{-}$ and CO \citep{Dawley2014}. 

Very recently, two independent investigations were made to study the reaction of H atoms on NH$_2$CHO ice \citep{Haupa2019, Suhasaria2020}. HNCO was formed through H-abstraction reactions via a H$_2$NCO radical intermediate when NH$_2$CHO was subjected to H atoms in the para-hydrogen quantum-solid matrix host (Haupa et al. 2019). NH$_2$CHO at 12 K exposed to H atoms was studied to estimate the effective destruction cross-section. This quantity was used to evaluate the destruction rate of NH$_2$CHO by H atoms in space and compared with the destructive effects of cosmic rays and UV photons. However, the absence of destruction cross-section for NH$_2$CHO under Ly$\alpha$ (10.2 eV) photons only allowed the estimation of an upper limit for the destruction rate under interstellar medium conditions. The present study reports the results of the interaction of Ly$\alpha$ with pure NH$_2$CHO ice film at 12 K. The NH$_2$CHO destruction and new products formed after UV irradiation have been studied by Fourier-transform infrared (FTIR) spectroscopy. The destruction cross-section for NH$_2$CHO and the cumulative formation cross-section for the formed new species have been obtained. The determination of the destruction cross-section has then allowed us to compare the destructive effects of H atoms, cosmic rays and UV photons on pure solid NH$_2$CHO in dense interstellar cloud conditions.

\section{Experimental methods}

Experiments were performed in a high vacuum chamber, which is described in detail in previous works \citep{Mennella2006b, Suhasaria2020}. The salient features of the set-up relevant to this work are briefly described here. The main chamber, with a base pressure better than 10$^{-8}$ mbar, is equipped with a closed-cycle helium cryostat and a dosing unit. A caesium iodide (CsI) substrate was mounted on the cold finger of the cryostat such that the substrate can be cooled down to 12 K. NH$_2$CHO (EMSURE grade; Merck) was purified by several freeze-pump-thaw and deposited on the substrate from the dosing unit with a pressure better than 10$^{-6}$ mbar. The ice film deposition is expected to be non-homogeneous.

To study the effects induced by UV photons, the NH$_2$CHO ice film was irradiated using a microwave excited hydrogen flow discharge lamp attached to the main chamber with a magnesium fluoride (MgF$_2$) window. Ly$\alpha$ (10.2 eV) emission accounts for 97\% of the total UV emission of the source \citep{Mennella2006b}. The UV beam forms an angle of 22.5$^{\circ}$ with the normal to the substrate. During sample irradiation, the UV flux (fluence) was monitored by measuring the current (charge) generated by photoelectric effect on a platinum wire inserted between the source and the substrate. Details on the calibration of the wire sensor and on the measure of the flux at the substrate position are given in \citet{Mennella2006b}.

The NH$_2$CHO ice film was studied in situ before and during irradiation in transmittance mode in the mid-infrared range at a resolution of 2 cm$^{-1}$ using a FTIR spectrometer (Bruker Vertex 80V) with a Mercury Cadmium Telluride (MCT) detector. For each single measurement, 1024 scans were co-added. The background was acquired for any spectral measurement of NH$_2$CHO before and after each irradiation step.

\begin{table*}
\caption{IR absorption band positions of pure NH$_2$CHO ice and the new species formed after the UV irradiation of NH$_2$CHO. Only the band strengths of those modes are reported whose column densities were evaluated.}              % title of Table
\label{table:1}      % is used to refer this table in the text
\centering                                      % used for centering table
\begin{tabular}{c c c c}          % centered columns (4 columns)
\hline\hline                        % inserts double horizontal lines
Molecule & Band position & Vibration mode & Band strength \\ 
& cm$^{-1}$ & & cm molecule$^{-1}$ \\% table heading
\hline                                   % inserts single horizontal line
NH$_2$CHO & 3355 & $\nu_{\mathrm{a}}$ (N-H) & $1.4\times10^{-16}$\tablefootmark{a} \\  % inserting body of the table
& 3170 & $\nu_{\mathrm{s}}$ (N-H) &  \\
& 2878 & $\nu$ (C-H) & $4.7\times10^{-18}$ \\
& 1690 &$\nu$ (C=O)& $6.5\times10^{-17}$ \\
& 1631 &$\delta$ (N-H) & \\
& 1386 &$\delta_{\mathrm{i}}$ (C-H)  & \\
& 1324 & $\nu$ (CN) &  \\
\hline
NH$_3$ & 3375 &$\nu$ (N-H) & \\
NH$_4^+$ & 3210 &$\nu$ (N-H) & \\
& 3065 &$\nu$ (N-H) & \\
& 1483 &$\delta$ (N-H) & \\
CO$_2$ & 2341 & $\nu_{\mathrm{a}}$ (C-O) &$7.6\times10^{-17}$\tablefootmark{b} \\
$^{13}$CO$_2$ & 2277 & $\nu_{\mathrm{a}}$ (C-O) & \\
HNCO & 2259 & $\nu_{\mathrm{a}}$ (N=C=O)&$7.2\times10^{-17}$\tablefootmark{c} \\
& 2240 & $\nu_{\mathrm{a}}$ (N=C=O)& \\
OCN$^{-}$ & 2163 & $\nu_{\mathrm{a}}$ (N=C=O) &$1.3\times10^{-16}$\tablefootmark{c} \\
CO & 2138 &$\nu$ (C$\equiv$O) &$1.1\times10^{-17}$\tablefootmark{d} \\
HCN & 2100 &$\nu$ (C$\equiv$N) &$1.1\times10^{-17}$\tablefootmark{e} \\
\hline                                             %inserts single line
\end{tabular}
\tablefoot{
\tablefoottext{a}{The band strength corresponds to ($\nu_{\mathrm{a}}$+$\nu_{\mathrm{s}}$). All the band strength values are taken from \citet{brucato2006b}.} \tablefoottext{b}{\citet{yamada1964}}
\tablefoottext{c}{\citet{van2004}
\tablefoottext{d}{\citet{jiang1975}}
\tablefoottext{e} {\citet{Gerakines2022}}}
}
\end{table*}

\begin{figure}
\begin{center}
\resizebox{\hsize}{!}{\includegraphics{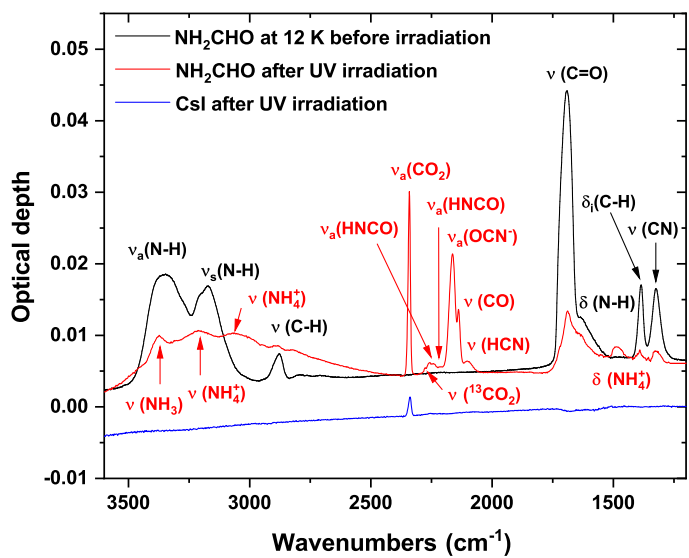}}
\caption{Mid-IR spectrum of a 29.5 ML thick NH$_2$CHO film at 12 K as deposited (black line) and after UV irradiation of 6.5$\times10^{18}$ photons cm$^{-2}$ (red line). Symbols $\nu$ and $\delta$ stand for stretching and bending modes, respectively. The subindex $\mathrm{a}$, $\mathrm{s}$, and $\mathrm{i}$ denotes antisymmetric, symmetric, and in-plane mode, respectively. The spectrum of a CsI substrate after UV irradiation of 6.5$\times10^{18}$ photons cm$^{-2}$ (blue line) is also shown for comparison. The red and blue curves are offset in ordinate by 0.0055 and -0.005, respectively, for the sake of clarity.}
   \label{fig1}
\end{center}
\end{figure}

\begin{figure*}
\begin{center}
\resizebox{\hsize}{!}{\includegraphics{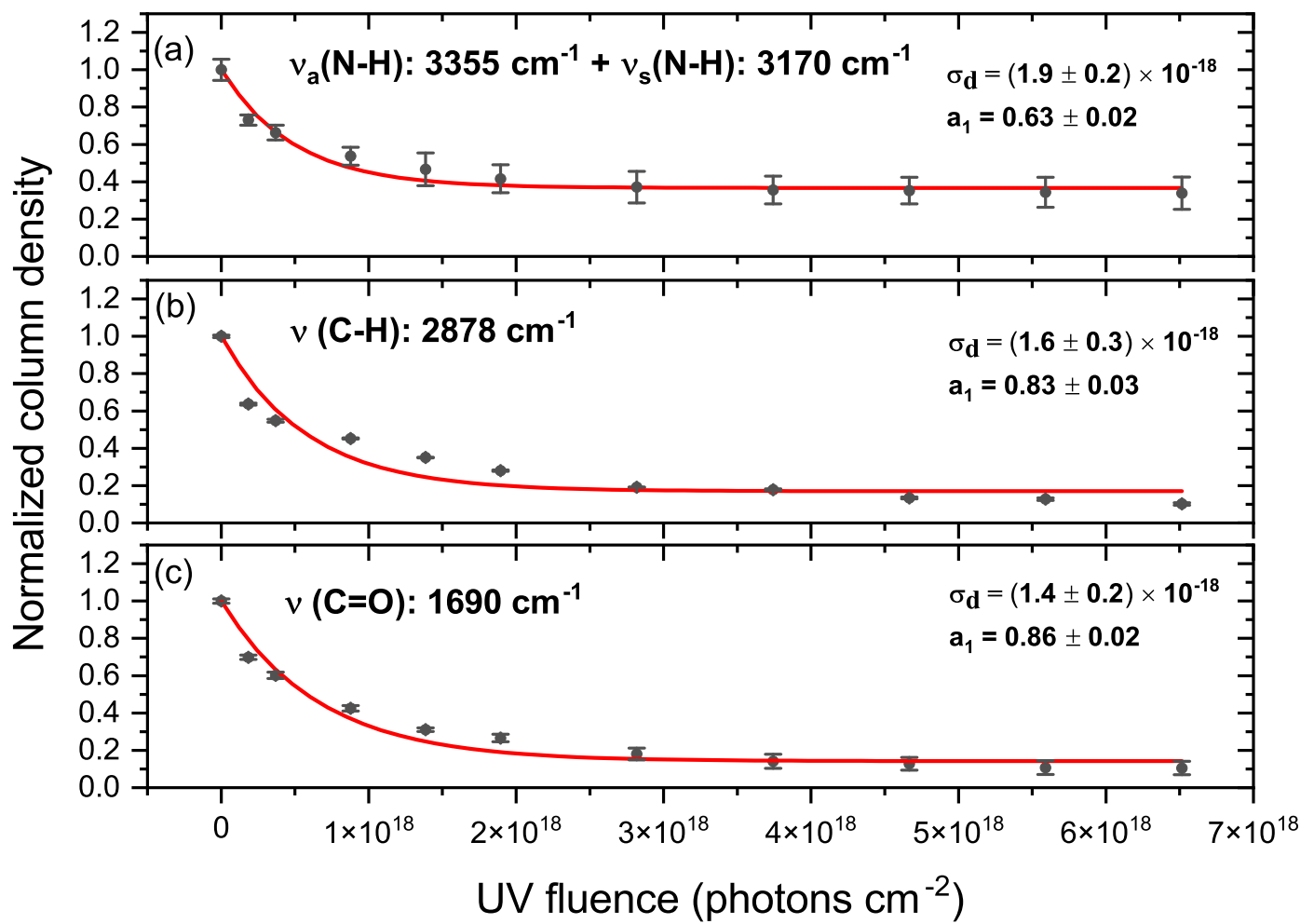}}
\caption{Evolution with UV photon fluence of the N-H (a), C-H (b) and C=O (c) normalised column density (filled circles). The normalisation factor is the corresponding initial column densities. The error arising due to baseline subtraction was taken into the error estimation. When the error bar is not visible, the error is within the size of the filled circle. The best fit to the data (red solid lines) is also shown. The estimated destruction cross-section ($\sigma_{d}$) and the asymptotic destruction ($\mathrm{a_1}$) for NH$_2$CHO are also reported.}
\label{fig2}
\end{center}
\end{figure*}

\begin{figure*}
\begin{center}
\resizebox{\hsize}{!}{\includegraphics{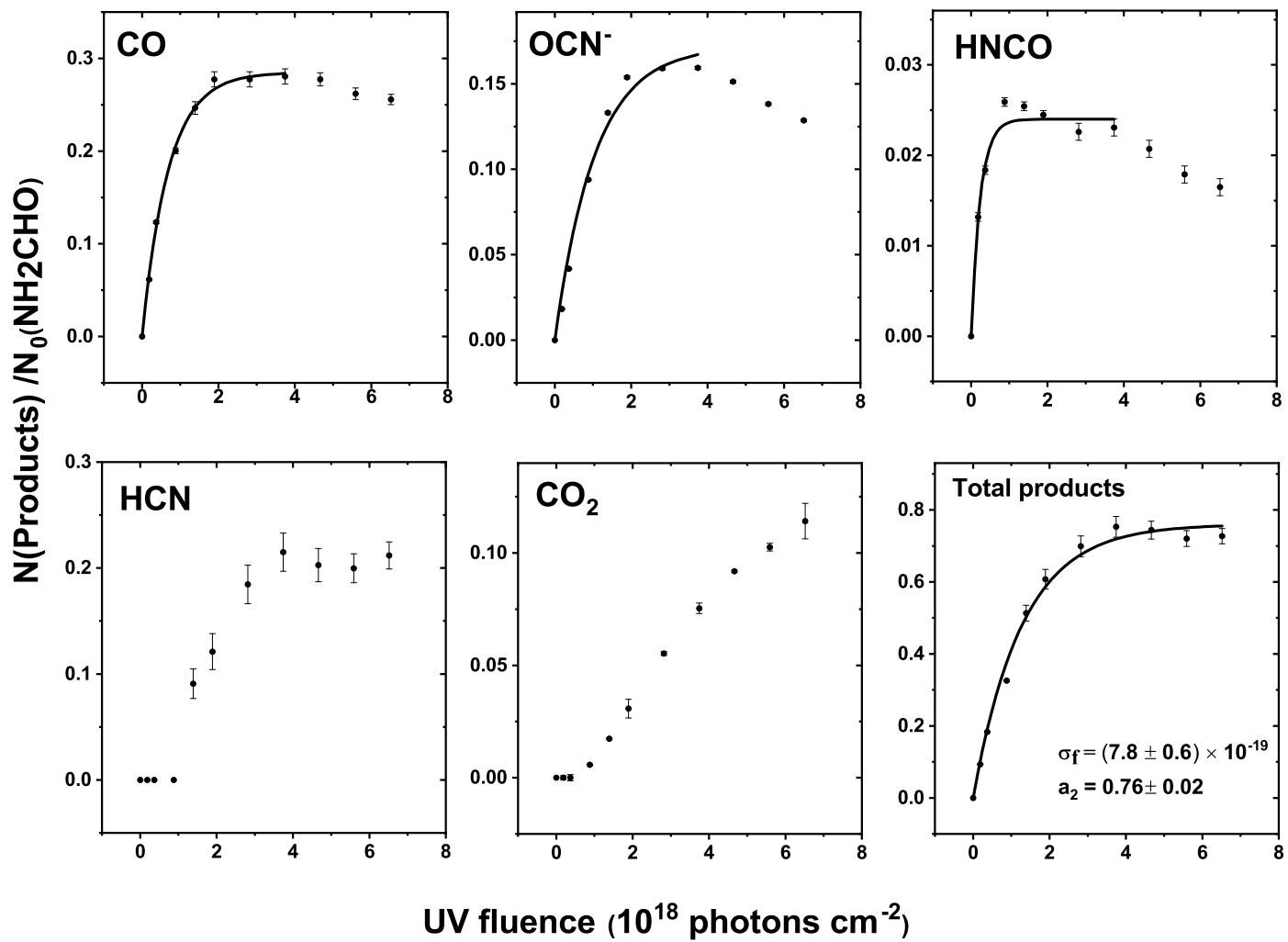}}
\caption{Evolution of normalised column density of different photo products with increasing UV photon fluence. The normalisation factor is the column density of NH$_2$CHO before UV irradiation. The IR absorption area is obtained by Gaussian fitting, with one standard deviation used as an error bar. When the error bar is not visible, the error is within the size of the symbol. The best fit to the cumulative abundance growth of the total photo product (solid black line) is also shown. The estimated formation cross-section ($\sigma_{f}$) and asymptotic formation ($\mathrm{a_2}$) are also reported. Similarly, the best fit to the growth curves (till F$_{UV}$ = 3.7$\times10^{18}$ photons cm$^{-2}$ ) of CO, OCN$^{-}$ and HNCO is also shown.}
   \label{fig3}
\end{center}
\end{figure*}

\begin{figure}
\begin{center}
\resizebox{\hsize}{!}{\includegraphics{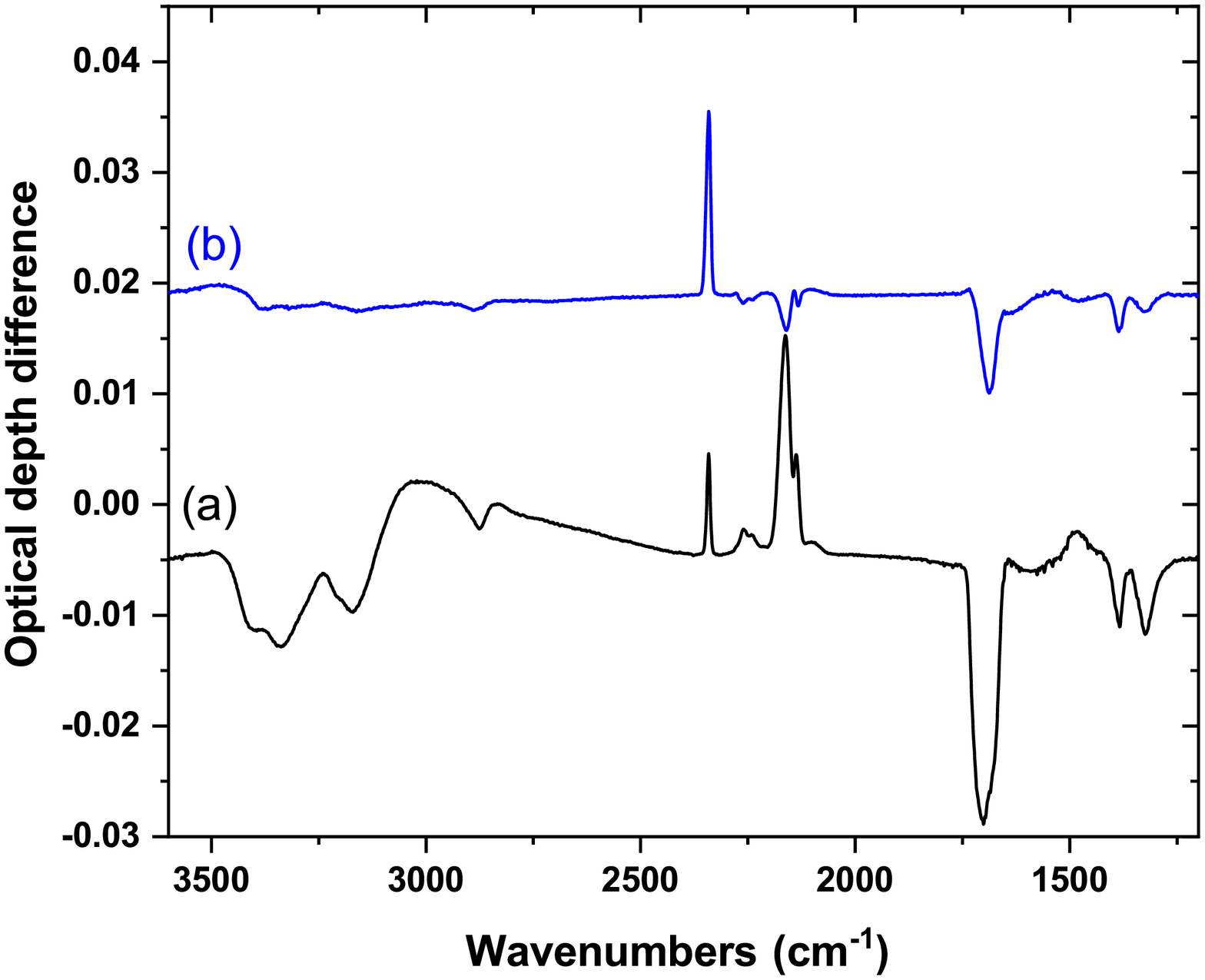}}
\caption{Difference IR spectrum of NH$_2$CHO after UV irradiation of: (a) 1.9$\times10^{18}$ photons cm$^{-2}$ with the ice as deposited; (b) 1.9$\times10^{18}$
 with  6.5$\times10^{18}$ photons cm$^{-2}$.}
   \label{fig4}
\end{center}
\end{figure}

\section{Results and Discussion}
\subsection{Ice growth}

The IR spectrum of pure solid NH$_2$CHO at 12 K is displayed in Fig.~\ref{fig1} and the corresponding fundamental vibration modes reported in Table~\ref{table:1} are in good agreement with previous results \citep{brucato2006b, Suhasaria2020}. The NH$_2$CHO ice film thickness was derived separately from the column density of combined symmetric and antisymmetric N-H, C-H and C=O stretching IR modes using the band strengths values listed in Table~\ref{table:1}. The average value, expressed in units of monolayer (1 ML = $10^{15}$ molecules cm$^{-2}$), is 29.5 ML which is equivalent to 0.03 $\mu$m. For the thickness determination, we took into account the cosine of the angle between the IR beam and the normal to the surface plane and the NH$_2$CHO ice density 0.937 g cm$^{-3}$ derived by \citet{brucato2006b}. The NH$_2$CHO ice thickness was found to be below the UV penetration depth since the mean free path of UV photons, d$_{UV}$, through the NH$_2$CHO films has been estimated to be 0.04 $\mu$m. The mean free path of UV photons was derived using the relation, d$_{UV}$ = 1/$\beta$ = 1/$\sigma$n where $\beta$ is the absorption coefficient, $\sigma$ is the UV photo-absorption cross-section of NH$_2$CHO and $n$ is the number of NH$_2$CHO molecules per unit volume. In the absence of direct measurements of the UV absorption cross-section for NH$_2$CHO ice, the gas phase value of $\textit{ca.}$ $2\times10^{-17}$ cm$^2$ \citep{gingell1997} acquired at 10.2 eV photons is taken into the calculation.

\subsection{UV irradiation of NH$_2$CHO}

Figure~\ref{fig1} also presents the IR spectrum of NH$_2$CHO ice after exposure to UV photon fluence of 6.5$\times10^{18}$ photons cm$^{-2}$. The intensity decrease of the NH$_2$CHO IR bands indicates that the NH$_2$CHO ice was depleted. At the same time, several new bands appear in the spectrum. The most intense band at 2341 cm$^{-1}$ band is due to the asymmetric stretch mode of CO$_2$. The second most intense band at 2163 cm$^{-1}$ shows a good agreement with the expected frequency of the N=C=O asymmetric stretching mode of OCN$^{-}$ \citep{Gerakines2004}. The counter ion NH$_4^+$ was clearly identified from the very broad N-H IR bending mode feature around 1483 cm$^{-1}$ \citep{raunier2004a}. Further evidence for the presence of NH$_4^+$ ion comes from the broad features at 3210 and 3065 cm$^{-1}$ overlapping with formamide N-H stretching modes. These peaks are in close agreement with the NH$_4^+$ peaks at 3206 and 3074 cm$^{-1}$ assigned previously by \citet{brucato2006b}.

At  2138 cm$^{-1}$ there is the symmetric stretch feature of CO which appears as a shoulder to the OCN$^{-}$ peak. The band observed at 2259 cm$^{-1}$ and a shoulder peak at the lower wavenumber side, centred around 2240 cm$^{-1}$, were assigned to the N=C=O asymmetric stretching mode of HNCO \citep{raunier2004b}. The shoulder peak at 2240 cm$^{-1}$ could also be associated to the N=N stretching mode of N$_2$O \citep{brucato2006b}. Furthermore, a shoulder peak at the higher wavenumber side of the 2259 cm$^{-1}$ band, centred around 2277 cm$^{-1}$, is attributed to $^{13}$CO$_2$ \citep{Modica2010}. Within the errors, the natural abundance ratio of $^{13}\mathrm{C}$:$^{12}\mathrm{C}$ = 0.011 is in agreement with the band area ratio of $^{13}\mathrm{CO_2}$:$^{12}\mathrm{CO_2}$ = 0.015. A peak around 2100 cm$^{-1}$ has been assigned previously to the C$\equiv$N stretching mode of HCN \citep{brucato2006b}. The 2100 cm$^{-1}$ band is the second strongest IR band of pure solid HCN ice after 3115 cm$^{-1}$ \citep{couturier2018} which has not been identified in the present experiment but could be obscured by the broad stretching band of N-H in NH$_2$CHO. NH$_3$ has also been tentatively identified after UV irradiation from the band present at 3375 cm$^{-1}$ due to N-H stretching based on the assignment by \citet{brucato2006b}. The band positions of the newly formed species are reported in Table~\ref{table:1}. A control experiment was also performed where a bare substrate was subjected to the same UV photon fluences as NH$_2$CHO ice. In this case, stretching mode of CO$_2$ was observed. At the maximum fluence, its intensity accounts for 12.5\% of the CO$_2$ band observed in the formamide irradiation (see Fig.~\ref{fig1}).

Figure~\ref{fig2} shows the evolution of normalised column densities,$\tau(t)$, of N-H, C-H and C=O stretching modes of NH$_2$CHO as a function of UV photon fluence, $\mathrm{F_{UV}}$. The experimental data was fitted following first order kinetic relation: 

\begin{equation}
\label{A}
\tau(t) = [\mathrm{1- a_1(1- e^{-\sigma_{d}F_{UV}})]}
\end{equation}

The equation refers to the destruction with the boundary condition that considers a residual reactant for $\mathrm{F_{UV}}$ = $\infty$. Fits to the experimental data in Fig.~\ref{fig2} allowed us to estimate the cross-sections and asymptotic values for the NH$_2$CHO destruction ($\mathrm{\sigma_{d}}$, $\mathrm{a_1}$).The destruction cross-section estimated from the three stretch modes are equal within the errors, suggesting that there is no difference in the chemical bond cleavage of corresponding functional groups upon irradiation by UV photons. This result differs from what we have found during H atoms exposure to NH$_2$CHO, where the C-H bonds are cleaved by H atom abstraction before the N-H bonds of the NH$_2$ group \citep{Suhasaria2020}. The estimated destruction cross-section was in agreement with the preliminary values obtained for formamide destruction by Ly$\alpha$ photons in an entirely different set-up (A. J. Escobar \& A. Ciaravella, 2022, priv. comm.). The average of the three estimated asymptotic fit parameters $\mathrm{a_{1, N-H}}$, $\mathrm{a_{1, CH}}$ and $\mathrm{a_{1, CO}}$ is 0.77 that corresponds to the destruction of 22.7 ML of the initial NH$_2$CHO ice. In the Ly$\alpha$ irradiation experiments of NH$_2$CHO deposited on a SiO$_2$ nanoparticles, after a UV photon fluence of 7$\times10^{19}$ photons cm$^{-2}$, $\textit{ca.}$ 29\%  of the initial NH$_2$CHO ice was destroyed \citep{Dawley2014b}. On the other hand, ion irradiation of NH$_2$CHO leads to the destruction of 64 and 78\% of the initial NH$_2$CHO molecules at fluences of 6.8$\times10^{14}$ and 1.4$\times10^{15}$ ions cm$^{-2}$, respectively \citep{brucato2006b}.

The evolution of column density of newly formed photo products normalised to the initial column density of NH$_2$CHO ice film as a function of UV fluence is shown in Fig.~\ref{fig3}. The band strength values used in the calculation of the individual column densities of the photo products are also listed in Table~\ref{table:1}. Some photo products displayed similar behaviour while others behave very differently with the increasing $\mathrm{F_{UV}}$. CO, OCN$^{-}$ and HNCO are formed immediately and rapidly at the beginning of the UV photolysis followed by slowing down and plateau within the error at UV fluence of about 4$\times10^{18}$ photons cm$^{-2}$. As the $\mathrm{F_{UV}}$ increases further, there is a decrease in the formation of all the three species. The formation of CO, OCN$^{-}$ and HNCO after UV irradiation of deposited NH$_2$CHO followed by a decrease in their intensity at the highest fluence is also clearly visible in the difference IR spectra in Fig.~\ref{fig4}. On the other hand, HCN shows a delayed formation with respect to CO,OCN$^{-}$ and HNCO. HCN appears only after UV photon fluence of 1.4$\times10^{18}$ photons cm$^{-2}$, but then the abundance increases rapidly and the normalised column densities are close to that of CO and OCN$^{-}$. Within the error, there is no apparent sign of the decrease in HCN formation at high fluences, as evident in Fig.~\ref{fig3} and Fig.~\ref{fig4}. In the case of CO$_2$, there is an initial phase of slow formation with increasing $\mathrm{F_{UV}}$ and only after UV fluence of 1.9$\times10^{18}$ photons cm$^{-2}$, CO$_2$ starts to increase rapidly. One can clearly see in the Fig.~\ref{fig4} that CO$_2$ intensity at the highest UV fluence increases with respect to that at 1.9$\times10^{18}$ photons cm$^{-2}$.

It is difficult to fit the abundance evolution of individual products over the entire UV fluence range by a single kinetic equation, due to the simultaneous formation and depletion behaviour. Therefore, following \citet{Chuang2021} a single first order kinetic relation was fitted to the cumulative abundance of all the photo products ($\chi (t)$): 

\begin{equation}
\label{B}
\chi(t) = \mathrm{a_2(1- e^{-\sigma_{f}F_{UV}})}
\end{equation}

where, $\sigma_{f}$ is the formation cross-section and $a_{2}$ is the asymptotic formation. Fits to the experimental data allowed us to estimate the effective formation cross-section, $\mathrm{\sigma_{f}}$ = $7.8\pm0.6 \times 10^{-19}$ cm$^{2}$ and the asymptotic value, $\mathrm{a_2}$ = $0.76\pm0.02$ for the products formation. The asymptotic value of total products formation with respect to the initial formamide column density exactly matches the average formamide destruction of 0.77. Of course, there could be other minor species produced in the irradiation experiment that have not clearly been identified by IR spectroscopy.

In the Ly$\alpha$ processing reactions of pure NH$_2$CHO ice deposited on SiO$_2$ nanopowder, cross-sections $\mathrm{\sigma_{f, CO}}$ = $3.9 \times 10^{-20}$ and $\mathrm{\sigma_{f, OCN-}}$ = $3.6 \times 10^{-20}$ cm$^{2}$ were estimated for the formation of CO and OCN$^{-}$, respectively \citep{Dawley2014}. Furthermore, in our previous study \citep{Suhasaria2020} we estimated the effective formation cross-section of HNCO due to H atoms exposure of formamide to be $\mathrm{\sigma_{f, HNCO}}$ = $4.4 \times 10^{-17}$ cm$^{2}$. Therefore, for the sake of comparison, we tried to fit only the individual growth curves of CO, OCN$^{-}$ and HNCO up to the UV fluence of 3.7$\times10^{18}$ photons cm$^{-2}$ using the same exponential equation as used for the cumulative growth. We derived the formation cross-sections of $\mathrm{\sigma_{f, CO}}$ = $1.5\pm0.1 \times 10^{-18}$, $\mathrm{\sigma_{f, OCN-}}$ = $1\pm0.1 \times 10^{-18}$ and $\mathrm{\sigma_{f, HNCO}}$ = $4.2\pm0.5\times 10^{-18}$ cm$^{2}$ for CO, OCN$^{-}$ and HNCO, respectively. The CO and OCN$^{-}$ formation cross-sections are two orders of magnitude higher than those estimated by \citet{Dawley2014}. The above two experiments differ in surface temperature and the surface type that impacts the formation cross-section. Out of the two, the primary impact would be of the surface temperature since there is an increase in the radical recombination efficiency with temperature as radicals diffuse faster within the ice. This would result in lower formation of CO and OCN$^{-}$ from NH$_2$CHO at higher surface temperature. On the other hand, the impact of SiO$_2$ nanoparticle surface could be less relevant since 600 ML thick ice was deposited on top of the surface for UV irradiation experiment. The HNCO formation cross-section that we derived in this experiment is an order of magnitude lower than that obtained previously by the exposure of H atoms on formamide \citep{Suhasaria2020}. The lower value of the HNCO formation cross-section can be taken as an indirect measure of the lesser stability of formamide under H atoms exposure compared to UV photons.

\begin{table*}
\caption{Destruction cross-sections of NH$_2$CHO under different energetic processing and the corresponding rates in the dense interstellar clouds.}
\label{table2}
\begin{tabular}{lcccc}
\hline
\hline
\multicolumn{1}{c}{Processing}& \multicolumn{1}{c}{Destruction cross-section}&\multicolumn{1}{c}{Flux}&\multicolumn{1}{c}{Destruction rate}\\
 &$\mathrm{\sigma_{d}}$, (cm$^2$) &$\mathrm{\Phi}$, (cm$^{-2}$ s$^{-1}$) & $\mathrm{R_{d}}$, (s$^{-1})$\\
\hline
UV photons & $1.9\pm0.2 \times 10^{-18}$ & $4.8 \times 10^{3}$\tablefootmark{c} & $9.1 \times 10^{-15}$\\
H atoms & $3.0\pm0.6 \times 10^{-17}$\tablefootmark{a} & $9.1 \times 10^{4}$\tablefootmark{b} & $2.7 \times 10^{-12}$ \\
Cosmic rays   & $3.7\pm0.4 \times 10^{-16}$\tablefootmark{d} & 1\tablefootmark{e} & $3.7 \times 10^{-16}$ \\

\hline
\end{tabular}
\tablefoot{
\tablefoottext{a}{\citet{Suhasaria2020}}
\tablefoottext{b}{\citet{Mennella2006a}
\tablefoottext{c} {corresponds to 10 eV photons \citep{Mennella2003}}
\tablefoottext{d} {corresponds to 1 MeV protons under monoenergetic approximation (G. A. Baratta \& M. E. Palumbo, 2020, priv. comm.)}
\tablefoottext{e} {\citet{Mennella2003}}}
}
\end{table*}

In Fig.~\ref{fig3} we see that growth curves of all the photo products except CO$_2$ resembles a first order kinetic behaviour, except for the very high fluences, which means they must be formed directly from NH$_2$CHO. CO$_2$ could have been formed by the effect of UV photons on CO and H$_2$O molecules formed in the reaction network \citep{watanabe2002}. This may explain why CO$_2$ formation rate is low at the initial fluences and increases only when more CO is produced in the ice mixture. At the highest UV fluence, there is not only a small decrease in the intensity of NH$_2$CHO but also of all the newly formed photo products except CO$_2$ and HCN. This suggests that in addition to NH$_2$CHO, there is processing of those species produced in the ice mixture at the highest fluence, hinting at a complex reaction network in play. It is beyond the scope of the present work to decipher such a complex reaction network and discuss every possible routes to molecule formation.

Despite the complexity in the reaction network, the following possible reaction pathways have been proposed based on the kinetic theory calculations by \citet{nguyen2011}. For the formation of CO, C-N bond in NH$_2$CHO should be dissociated and could have proceeded either in single step or energetically favourable two step process involving aminohydroxy carbene intermediate (NH$_2$-\"C-OH). We have tentatively identified NH$_3$ in the formamide spectrum after UV irradiation as discussed above.

\begin{equation}
\label{reaction1}
\mathrm{NH_2CHO} \xrightarrow{\mathrm{h\nu}} \mathrm{NH_2\ddot{C}OH} \rightarrow \mathrm{CO} + \mathrm{NH_3}
\end{equation}

HNCO should be formed by H$_2$ loss and proceeds either through energetically favourable single step 1,2-H$_2$ elimination or a two step process.

\begin{equation}
\label{reaction2}
\mathrm{NH_2CHO} \xrightarrow{\mathrm{h\nu}} \mathrm{HNCO} + \mathrm{H_2}
\end{equation}

HCN can be formed due to the H$_2$O loss from NH$_2$CHO. The delayed formation of HCN seems to be in agreement with a multi-step formation process via a formimidic acid (HN=CH-OH) conformer as proposed by \citet{nguyen2011}.

\begin{equation}
\label{reaction3}
\mathrm{NH_2CHO} \xrightarrow{\mathrm{h\nu}} \mathrm{HNCHOH} \rightarrow \mathrm{HCN} + \mathrm{H_2O}
\end{equation}

Three specific routes were suggested for the formation of OCN$^{-}$ in the condensed phase from the UV irradiation of NH$_2$CHO ice \citep{Dawley2014}. First is a direct photodissociation followed by ionisation reaction, second is direct ionisation followed by ion-electron recombination and electron capture, and third is a direct dissociative electron attachment (preceded by a direct excitation in formamide, indicated by an asterix).

\begin{equation}
\label{reaction4}
\mathrm{NH_2CHO} \xrightarrow{\mathrm{h\nu}} \mathrm{OCN^-} + \mathrm{H_2} + \mathrm{H^+}
\end{equation}

\begin{equation}
\label{reaction5}
\mathrm{NH_2CHO} \xrightarrow{\mathrm{h\nu}} \mathrm{NH_2CHO^{.+}} \rightarrow \mathrm{OCN^-} + \mathrm{H_2}
\end{equation}

\begin{equation}
\label{reaction6}
\mathrm{e^-} + \mathrm{NH_2CHO} \rightarrow (\mathrm{HNCHOH})^{*+} \rightarrow \mathrm{OCN^-} + \mathrm{H_2} + \mathrm{H}
\end{equation}

Since we can not rule out any possibilities, any of the three or all three reactions could potentially yield OCN$^{-}$ in our study.

\section{Astrophysical implications}

Solid state abundances of NH$_2$CHO molecule can be predicted from gas-grain chemical models only when a complete picture of its formation and destruction under different conditions are taken into account. The knowledge of the corresponding rates, related to the cross-sections, is therefore necessary. Formamide ice could most likely be present in the dense interstellar cloud conditions not in pure form but mixed with water or other interstellar ice components. However, formamide is more refractory than water or other volatiles which means small quantities of pure formamide ice could exist in elevated grain temperatures. In addition, to determine the extent of destruction under various energetic processing agents and to compare their effects we have considered here a pure ice as previously done. \citet{Dawley2014} showed that H$_2$O plays a catalytic role by increasing the product formation when H$_2$O mixed NH$_2$CHO ice is exposed to Ly$\alpha$ photons. However, the impact of other volatiles is still unknown and would require dedicated experiments to gain further insight.

In fact, in our previous study we derived the effective destruction cross-section, $\mathrm{\sigma_{d, H}}$ = $3.0\pm0.6 \times 10^{-17}$ cm$^2$ for pure NH$_2$CHO due to thermal H atoms exposure. This was found to be an order of magnitude lower than that derived for the destruction of formamide by 200 keV H$^+$, simulating the effects of cosmic rays. Under the approximation of monoenergetic 1 MeV protons, a value of $\mathrm{\sigma_{d, 1 MeV}}$ = $3.7\pm0.4 \times 10^{-16}$ cm$^2$, was derived (G. A. Baratta \& M. E. Palumbo, 2020, priv. comm.). Furthermore, due to the lack of experimentally derived NH$_2$CHO destruction cross-section under Ly$\alpha$ (10.2 eV) photons, we made an assumption to derive only an upper limit value of the destruction cross-section, $\mathrm{\sigma_{UV}}$ = $7.5 \times 10^{-16}$ cm$^2$. We argued that high energy protons would induce multiple bond breaking in the molecules along the “hot track”\footnote{The path of the swift ions through the ice film which gets locally heated from the energy of the ions.} compared to a single photolysis step by Ly$\alpha$ photons and therefore the UV destruction cross-section should be lower than that for energetic protons. In agreement with the above argument, we found that UV destruction cross-section of formamide, $\mathrm{\sigma_{d, UV}}$ = $1.9\pm0.2\times 10^{-18}$ cm$^2$, estimated in the present work is two orders of magnitude lower than that obtained for energetic protons. Moreover, $\mathrm{\sigma_{d, UV}}$ is about ten times lower than $\mathrm{\sigma_{d, H}}$, the destruction cross-section by H atoms (see Table~\ref{table2}).

The derivation of the cross-section further allowed to evaluate the NH$_2$CHO destruction rate under UV photons in dense clouds and compare it with the destruction rates induced by H atoms and cosmic rays. In the dense cloud cores, formamide ice should be shielded from the external UV radiation but there are locally produced UV rays resulting from cosmic-rays induced ionisation of hydrogen. The energy of the UV photons that impinges on the interior of dense clouds resembles the Ly$\alpha$. Taking into account the UV photons flux of $4.8 \times 10^{3}$ cm$^{-2}$ s$^{-1}$ \citep{Mennella2003} in those environments, the destruction rate, $\mathrm{R_{d}}$ = $9.1 \times 10^{-15}$ s$^{-1}$ was obtained. Although this rate was found to be an order of magnitude higher than the cosmic rays, the rate was three orders of magnitude lower than that induced by H atoms. This means that H atoms induce the higher destruction efficiency in formamide compared to UV photons or cosmic rays in those environments. The destruction rates for formamide under different energetic processing are also tabulated in Table~\ref{table2}.

\section{Conclusions}

This experimental study of Ly$\alpha$ irradiation of NH$_2$CHO ice at 12 K under high vacuum conditions was intended to understand its photo stability under dense cloud conditions. The UV photolysis results in the formation of new products CO, NH$_4^+$ OCN$^-$, HCN, HNCO, and CO$_2$ which were identified by FTIR spectroscopy. The formation mechanism of other photo products were also discussed. The destruction of N-H, C-H and C=O functional groups in NH$_2$CHO occurred in a single step unlike the case of H atoms bombardment as examined in our previous study. For the first time, the Ly$\alpha$ destruction cross-section of NH$_2$CHO and the cumulative formation cross-section of different photo products are estimated. The comparison of the destruction rate of NH$_2$CHO in dense clouds obtained in the present work with those induced by other processing indicates that H atoms interaction remains the driving mechanism for the destruction of this molecule in those environments.

\begin{acknowledgements}

This work has been supported by the project PRIN-INAF 2016 "The Cradle of Life- GENESIS- SKA" (General Conditions in Early Planetary Systems for the rise of life with SKA). The authors wish to thank Angela Ciaravella and Antonio Jimenez Escobar for fruitful discussions.

\end{acknowledgements}

\bibliographystyle{aa}
\bibliography{Reference}
\end{document}